\documentclass[
showkeys,
reprint,
superscriptaddress,
nofootinbib,
nobibnotes,
longbibliography,
aps,
prb,
floatfix,
nofootinbib
]{revtex4-2}

\usepackage{soul}
\usepackage{amsmath}
\usepackage{amssymb}
\usepackage{amsfonts}
\usepackage{tablefootnote}
\usepackage{graphicx}
\usepackage{color}
\usepackage[dvipsnames]{xcolor}
\usepackage{bbm}
\usepackage[breaklinks=true,colorlinks=true,linkcolor=blue,urlcolor=blue,citecolor=blue]{hyperref}
\usepackage{array}
\usepackage{booktabs}
\usepackage{mathptmx}
\usepackage{placeins}

\DeclareMathAlphabet{\mathcal}{OMS}{cmsy}{m}{n}
\newcolumntype{P}[1]{>{\centering\arraybackslash}p{#1}}
\newcolumntype{M}[1]{>{\centering\arraybackslash}m{#1}}

\bibstyle{Apsrev4-2}

\begin{document}

\title{Annealing-enhanced spin–orbit effects in non-centrosymmetric superconducting NbRe films}


\author{Zahra Makhdoumi Kakhaki}
\email{z.makhdoumi-kakhaki@tu-braunschweig.de}
    \affiliation{Cryogenic Quantum Electronics, Institute for Electrical Measurement Science and Fundamental Electrical Engineering (EMG) and Laboratory for Emerging Nanometrology (LENA), Technische Universit\"at Braunschweig, 38106 Braunschweig, Germany}

\author{Yuriy Yerin}
    \affiliation{Istituto di Struttura della Materia of the National Research Council, via Salaria Km 29.3, I-00016 Monterotondo Stazione, Italy}
    \affiliation{Cryogenic Quantum Electronics, Institute for Electrical Measurement Science and Fundamental Electrical Engineering (EMG) and Laboratory for Emerging Nanometrology (LENA), Technische Universit\"at Braunschweig, 38106 Braunschweig, Germany}

\author{Francesco Avitabile}
    \affiliation{Dipartimento di Fisica “E. R. Caianiello”, Università degli Studi di Salerno, I-84084 Fisciano, Salerno, Italy}
    
\author{Abhishek Kumar}
     \affiliation{Dipartimento di Fisica “E. R. Caianiello”, Università degli Studi di Salerno, I-84084 Fisciano, Salerno, Italy}
     \affiliation{CNR-SPIN, c/o Università degli Studi di Salerno, I-84084 Fisciano, Salerno, Italy}

\author{Francesco Colangelo}
       \affiliation{Dipartimento di Fisica “E. R. Caianiello”, Università degli Studi di Salerno, I-84084 Fisciano, Salerno, Italy}
       \affiliation{CNR-SPIN, c/o Università degli Studi di Salerno, I-84084 Fisciano, Salerno, Italy}
       
\author{Carla Cirillo}
       \affiliation{CNR-SPIN, c/o Università degli Studi di Salerno, I-84084 Fisciano, Salerno, Italy}

\author{Oleksandr V. Dobrovolskiy}
    \affiliation{Cryogenic Quantum Electronics, Institute for Electrical Measurement Science and Fundamental Electrical Engineering (EMG) and Laboratory for Emerging Nanometrology (LENA), Technische Universit\"at Braunschweig, 38106 Braunschweig, Germany}
    \affiliation{FLUXONICS---The European Foundry for Superconducting Electronics e.V., 38116 Braunschweig, Germany}

\author{Carmine Attanasio}
       \affiliation{Dipartimento di Fisica “E. R. Caianiello”, Università degli Studi di Salerno, I-84084 Fisciano, Salerno, Italy}
       \affiliation{CNR-SPIN, c/o Università degli Studi di Salerno, I-84084 Fisciano, Salerno, Italy}
        \affiliation{Centro NANOMATES, c/o Università degli Studi di Salerno, I-84084 Fisciano, Salerno, Italy}

\begin{abstract}
$\text{Nb}_{0.18}\text{Re}_{0.82}$ (NbRe) is a non-centrosymmetric superconductor with a transition temperature $T_\mathrm{c}$ reaching $9\text{ K}$ in bulk form. While bulk and single-crystalline NbRe exhibit signatures of multigap superconductivity, thin films generally display a single-gap superconducting state due to structural disorder and reduced crystallite dimensions. Here, we investigate the impact of thermal annealing on the superconducting and normal-state magnetotransport properties of NbRe films. The temperature dependence of the upper critical field, $B_{\mathrm{c2}}(T)$, is analyzed within the microscopic Werthamer--Helfand--Hohenberg (WHH) framework, while the normal-state magnetoconductivity is described using the three-dimensional Kawabata weak-localization/weak-anti-localization model. Annealing drives a pronounced change in the electronic response, manifested by a strong weak anti-localization behavior in the normal state and an upper critical field that surpasses both the conventional orbital-limiting field and the Pauli paramagnetic limit. The microscopic analysis reveals a strong intrinsic increase in the relative spin--orbit scattering strength, with the annealed film showing a significantly enhanced spin--orbit-to-dephasing field ratio. These findings provide direct, independent evidence that thermal modification of the NbRe microstructure successfully amplifies spin--orbit-mediated quantum transport, which acts as the key mechanism protecting the non-centrosymmetric superconducting state against paramagnetic pair-breaking well beyond conventional theoretical boundaries.
\end{abstract}

\maketitle

\section{Introduction}
Conventional superconductivity (SC) is well described by the Bardeen--Cooper--Schrieffer (BCS) theory, in which Cooper pairs form spin-singlet states with opposite spin and momentum. In contrast, spin-triplet SC involves pairing of electrons with parallel spins \cite{bauer2012non}. Despite extensive efforts, unambiguous evidence for intrinsic spin-triplet SC remains elusive. Many studies have therefore focused on engineering triplet correlations in hybrid structures comprising conventional SCs and magnetic or spin--orbit-active materials, where equal-spin Cooper pairs can be generated via proximity effects \cite{khaire2010observation,robinson2010controlled}. Such systems are of particular interest for superconducting spintronics \cite{Linder2015}, where dissipationless spin-polarized supercurrents offer a route toward novel device concepts and energy-efficient information processing \cite{eschrig2011spin}.

Non-centrosymmetric superconductors (NCSs) provide a natural platform for exploring unconventional pairing states. In these materials, the absence of inversion symmetry induces antisymmetric spin--orbit coupling (ASOC), lifting spin degeneracy and enabling the admixture of even-parity spin-singlet and odd-parity spin-triplet components in the SC order parameter \cite{bauer2012non,Yip_review,Smidman2017}. Among NCSs, the binary alloy Nb$_{0.18}$Re$_{0.82}$ (NbRe) has emerged as a promising platform for exploring unconventional SC. NbRe crystallizes in the $\alpha$-Mn structure (space group $I\overline{4}3m$) and exhibits a relatively high critical temperature of about $9$\,K \cite{Cirillo2015,Karki2011,Chen2013,sundar2019study}.

Experimental investigations of NbRe have revealed SC behavior that depends strongly on sample morphology. In bulk NbRe single crystals, spectroscopic measurements have reported signatures of two SC gaps, suggesting a complex NCS state with possible unconventional pairing \cite{Cirillo2015}. By contrast, polycrystalline NbRe in both bulk and thin-film forms generally exhibits single-gap and conventional SC \cite{Cirillo2016,Karki2011,Chen2013}. Nevertheless, several experiments have indicated the possible presence of spin-triplet correlations in NbRe. Muon-spin rotation measurements revealed a spontaneous internal magnetic field below the SC transition temperature, suggesting time-reversal symmetry breaking \cite{ShangPRL2018}. Furthermore, ferromagnetic resonance measurements in NbRe/Co bilayers identified signatures of equal-spin triplet correlations through spin-pumping effects \cite{cirillo2023spin}, while recent NbRe-based SC spin-valve devices provided further evidence for their formation \cite{colangelo2025unveiling}.

Beyond its fundamental interest, NbRe exhibits properties attractive for applications. NbRe thin films and NbRe-based hybrid structures display ultrafast vortex dynamics and strong spin--orbit-related effects \cite{caputo2017nbre,de2026fast,kakhaki2024characterization,Cirillo2016}, making them promising for SC spintronics, fluxonic devices, and quantum technologies \cite{Smidman2017,Linder2015,Dob25nan,Dob26sst}. In addition, disordered NbRe films have shown excellent performance as SC single-photon detectors and emerged as candidates for gate-tunable SC devices and high-kinetic-inductance superinductors \cite{cirillo2020superconducting,caputo2017nbre,ejrnaes2022single,cirillo2024single,koch2024gate,battisti2024demonstration}.

The SC properties of NbRe thin films are highly sensitive to microstructure, including grain size, crystallographic texture, and surface conditions \cite{cirillo2022polycrystalline,makhdoumi2024effect}. Recent synchrotron spectroscopy studies have shown that annealing promotes the emergence of the non-centrosymmetric $\chi$ phase, which is absent in as-grown films \cite{makhdoumi2024effect}. These observations suggest that annealing may restore intrinsic non-centrosymmetric SC characteristics that are suppressed in disordered polycrystalline films by grain-boundary scattering. Consequently, annealing-induced crystallization provides a route to access SC properties closer to those of single crystals and to tune the SC response through microstructural control \cite{Cirillo2015}.

Here, we investigate the influence of annealing on the SC and normal magnetotransport properties of NbRe thin films through measurements of the temperature dependence of the upper critical field, $B_{\mathrm{c2}}(T)$, and magnetoconductivity (MC). The $B_{\mathrm{c2}}(T)$ data are analyzed within the Werthamer--Helfand--Hohenberg (WHH) framework~\cite{werthamer1966temperature}, while the normal-state MC is described using the three-dimensional Kawabata weak-localization/weak-anti-localization (WL/WAL) model. We find that annealing strongly enhances spin--orbit-driven effects, leading to pronounced WAL behavior and an upper critical field exceeding the conventional Pauli paramagnetic limit. The Kawabata analysis further reveals a substantial increase in spin--orbit scattering strength. These results establish annealing as an effective means of controlling spin--orbit-mediated transport and SC behavior in NbRe thin films.

\section{Superconducting properties}

\subsection{Samples}

NbRe thin films with thicknesses of $8$, $20$, and $150\,\mathrm{nm}$ were deposited by dc magnetron sputtering from a Nb$_{0.18}$Re$_{0.82}$ target onto Si/SiO$_2$ substrates at room temperature. The base pressure prior to deposition was in the low $\sim10^{-8}$\,Torr regime, and sputtering was performed in an Ar atmosphere at a pressure of $3$\,mTorr. After deposition, one 20\,nm-thick film was left not annealed, while two other 20\,nm-thick films were annealed at $600^\circ$C for 30 or 45\,min, followed by a second annealing step at $300^\circ$C for 30\,min. The 150\,nm-thick film was annealed at $600^\circ$C for 6\,h and subsequently at $300^\circ$C for 1\,h. The 8\,nm-thick film was not annealed and served as a reference sample. All annealing treatments were performed in flowing argon to minimize oxidation as described in Ref. ~\cite{makhdoumi2024effect}. Throughout the manuscript, as-grown films are denoted as $\text{A}d$, where $d$ represents the thickness in nanometers (i.e., A8, A20, and A150). For the annealed samples, a distinction in nomenclature is adopted based on the thickness series. Specifically, for the $20\text{ nm}$ films where multiple annealing durations were explored, samples are designated as $\text{N}20\text{-}\tau$, where $\tau$ denotes the duration of the first annealing step (namely, N20-30min and N20-45min). Conversely, for the $150\text{ nm}$ thickness where only a single annealing procedure was investigated, the sample is simply labeled as N150.

\subsection{Structural characterization}

\begin{figure}[t!]
  \centering
  \includegraphics[width=8.6cm]{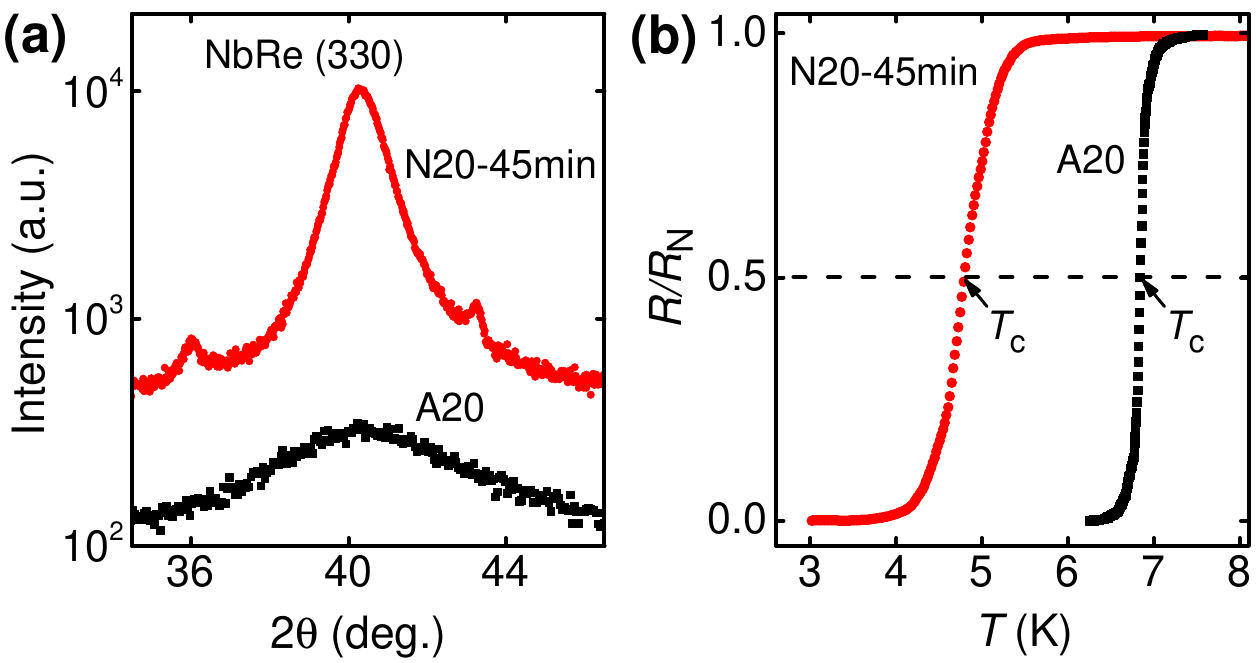}
  \caption{\label{R-T-XRD} 
  (a) XRD patterns of the as-grown (A20) and annealed (N20-45min) 20\,nm-thick NbRe films. (b) Temperature dependence of the normalized electrical resistance for the same samples.}
\end{figure}

The structural properties of the NbRe thin films were investigated by X-ray diffraction (XRD) using a Cu-anode diffractometer operating with characteristic $\mathrm{Cu}K_{\alpha}$ ($\lambda_{\alpha}=1.5406$\,\AA) radiation. Figure~\ref{R-T-XRD}(a) presents the $\theta$--$2\theta$ diffraction patterns of the $20$\,nm-thick as-grown (A20) and annealed (N20-45min) films in the vicinity of the principal NbRe reflection. Both samples exhibit a single diffraction peak indexed to the $(330)$ planes of the non-centrosymmetric $\alpha$-Mn NbRe phase (space group $I\bar{4}3m$). Upon annealing, the diffraction peak shifts from $2\theta \approx 40.82^\circ$ for A20 to $2\theta \approx 40.28^\circ$ for N20-45min and becomes sharper, indicating crystallite growth and improved crystalline order.

The lattice parameter was determined from the angular position of the $\mathrm{NbRe}(330)$ reflection using the cubic relation,
\[
a=\frac{\sqrt{h^2+k^2+l^2}\,\lambda}{2\sin\theta},
\]
where $h=k=3$ and $l=0$ are the Miller indices of the reflecting planes. The calculated lattice parameter increases from $a=9.37$\,\AA\ for A20 to $a=9.49$\,\AA\ for N20-45min, consistent with a slight lattice expansion following annealing. This behavior is attributed to the relaxation of residual compressive stresses developed during sputter deposition together with local atomic rearrangement toward a more equilibrium crystal structure. These lattice parameters, together with the increase in crystallite size from approximately $2$\,nm to $10$\,nm after annealing, are in good agreement with our previous studies on NbRe thin films~\cite{makhdoumi2024effect,cirillo2022polycrystalline}.

The average crystallite size $t_\mathrm{cryst}$ was estimated using the Scherrer relation, $t_\mathrm{cryst} = 0.9\lambda /(B_\mathrm{w}\cos\theta)$, where $B_\mathrm{w}$ is the full width at half maximum of the diffraction peak. The analysis reveals an increase in crystallite size from $\sim2$\,nm in the as-grown films to $\sim8$\,nm (N20-30min), $\sim10$\,nm (N20-45min), and $\sim15$\,nm (N150) after annealing~\cite{makhdoumi2024effect}. This pronounced grain growth highlights the strong impact of thermal treatment on the microstructural evolution of NbRe thin films~\cite{makhdoumi2024effect,cirillo2022polycrystalline}.

\subsection{Electrical resistance measurements}

Electrical transport measurements were performed in a closed-cycle cryostat equipped with a $7$\,T SC magnet. Electrical resistance measurements were done in the standard four-probe geometry using silver paste contacts. The temperature dependence of the resistance, $R(T)$, was measured in the range of 2--300\,K under applied perpendicular magnetic fields up to 6\,T. The SC transition temperature, $T_\mathrm{c}$, was defined using the 50\% criterion of the normal-state resistance, as indicated in Fig.~\ref{R-T-XRD}(b). 

Figure~\ref{R-T-XRD}(b) shows the temperature dependence of the normalized resistance, $R/R_\mathrm{N}$, where $R_\mathrm{N}$ denotes the normal-state resistance measured at $8$\,K, for the as-grown and annealed $20$\,nm-thick films. The superconducting transition temperature decreases from $T_\mathrm{c}=6.6$\,K for A20 to $T_\mathrm{c}=4.8$\,K for N20-45min. This annealing-induced suppression of $T_\mathrm{c}$ is consistent with previous observations in NbRe thin films~\cite{makhdoumi2024effect}. We attribute the reduction of $T_\mathrm{c}$ to the combined effects of residual oxidation during annealing, interfacial interdiffusion, and annealing-induced microstructural evolution accompanying the pronounced crystallite growth~\cite{grigorov1993interdiffusion,piallat2016investigation}.

\begin{figure*}[t]
    \centering
    \includegraphics[width=0.95\textwidth]{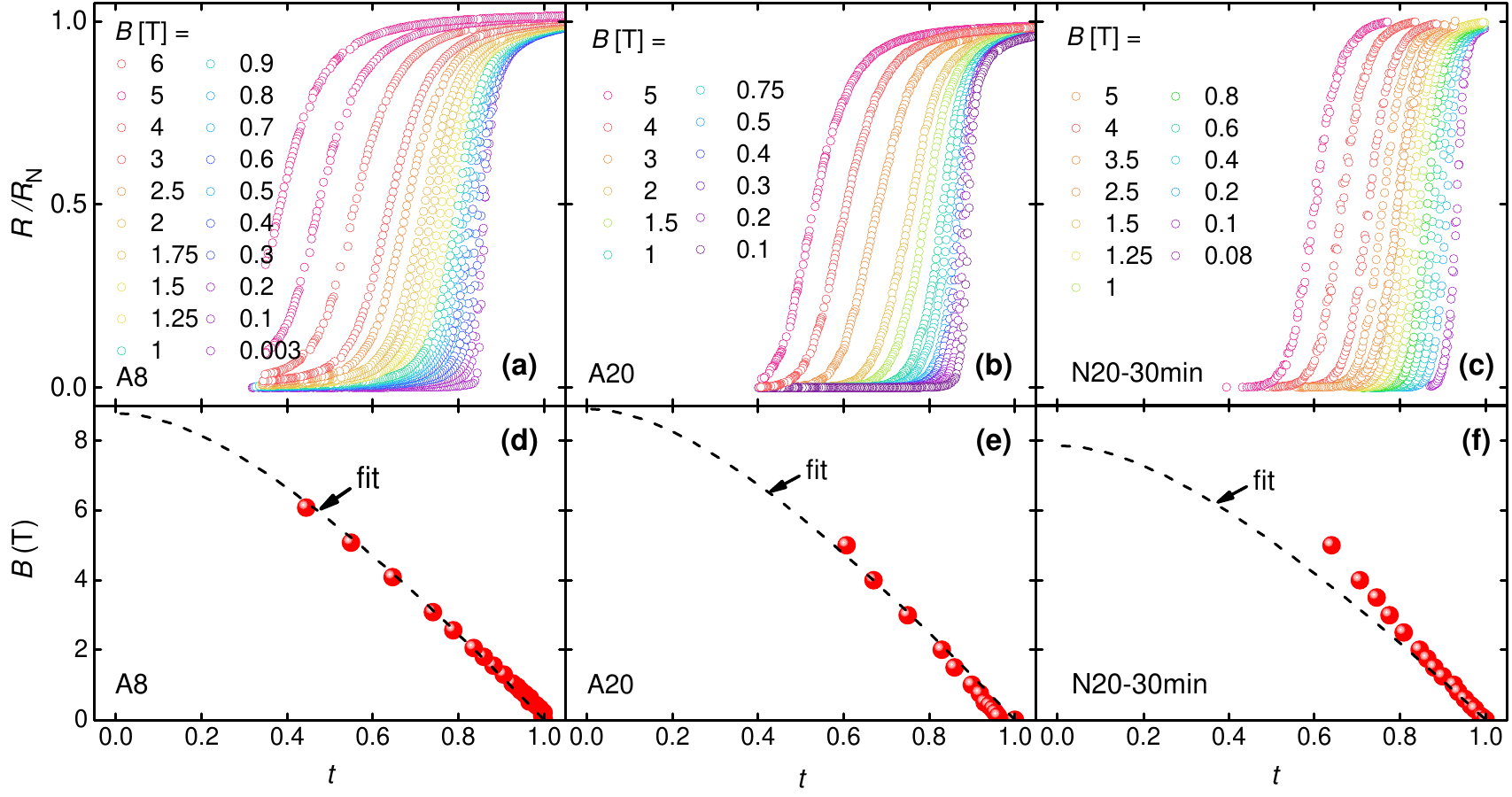}
    \caption{\label{transitions}
    (a--c) Normalized resistance as a function of the reduced temperature,
    $t=T/T_\mathrm{c}$, for NbRe films A8, A20, and N20-30min under
    different applied magnetic fields. (d--f) Upper critical field as a
    function of the reduced temperature for the corresponding films. The
    dashed curves represent fits to the WHH model, Eq.~(\ref{eq:WHH}),
    with $\alpha=0$ and $\lambda_{\mathrm{so}}=0$.}
\end{figure*}

\subsection{Upper critical field}

Figures~\ref{transitions}(a--c) show the superconducting transitions of films A8, A20, and N20-30min, represented by the normalized resistance $R/R_\mathrm{N}$ as a function of the reduced temperature $t = T/T_\mathrm{c}$ under various perpendicular applied magnetic fields. With increasing magnetic field, the resistive transitions systematically broaden and shift to lower temperatures. 

To investigate the dominant pair-breaking mechanisms and assess the role of spin--orbit effects, the $B_{\mathrm{c2}}(T)$ data extracted from the 50\% criterion were analyzed within the microscopic Werthamer--Helfand--Hohenberg (WHH) model~\cite{werthamer1966temperature,kakhaki2023upper}:
\begin{equation}
    \label{eq:WHH}
    \ln\left( \frac{1}{t} \right) = \sum_{\nu=-\infty}^{\infty} \left( \frac{1}{n} - \left[ n + \frac{\bar{h}}{t} + \frac{(\alpha\bar{h}/t)^2}{n + (\bar{h} + \lambda_{\mathrm{so}})/t} \right]^{-1} \right)
\end{equation}
where $n = |2\nu + 1|$, $\lambda_{\mathrm{so}} = \hbar / (3\pi k_\mathrm{B} T_\mathrm{c} \tau_{\mathrm{so}})$ is the spin--orbit scattering parameter, and $\bar{h} = 4B_{\mathrm{c2}}(T) \left[ \pi^2 T_\mathrm{c} \left| \frac{dB_{\mathrm{c2}}}{dT} \right|_{T=T_\mathrm{c}} \right]^{-1}$ represents the reduced dimensionless magnetic field. The Maki parameter $\alpha = \sqrt{2}B_{\mathrm{c2}}^{\mathrm{orb}}(0)/B_\mathrm{p}(0)$ quantifies the relative strength of Pauli paramagnetic limiting compared to orbital pair breaking~\cite{Maki1964}. Here, the orbital-limited upper critical field in the dirty limit is given by $B_{\mathrm{c2}}^{\mathrm{orb}}(0) = 0.69 T_\mathrm{c} \left| \frac{dB_{\mathrm{c2}}}{dT} \right|_{T=T_\mathrm{c}}$, while the Pauli paramagnetic limit is $B_p(0) = 1.84 T_\mathrm{c}$. 

\begin{table}[b]
\caption{\label{tab:parameters} 
Parameters extracted from the upper critical field analysis for the as-grown (A8, A20) and annealed (N20-30min) $\text{Nb}_{0.18}\text{Re}_{0.82}$ thin films.}
\begin{ruledtabular}
\begin{tabular}{lcccc}
Sample & A8 & A20 & N20-30min \\
\hline
$T_{\mathrm{c}}$ (K) & 5.32 & 6.60 & 6.30 \\
$D$ ($\text{cm}^{2}/\text{s}$) & 0.49 & 0.55 & 0.57 \\
$\xi(0)$ (nm) & 6.2 & 6.0 & 6.3 \\
$-\mathrm{d}B_{\mathrm{c2}}/\mathrm{d}T|_{T_{\mathrm{c}}}$ (T/K) & 2.23 & 1.99 & 1.96 \\
$B_{\mathrm{p}}(0)$ (T) & 9.8 & 12.1 & 11.6 \\
$B_{\mathrm{c2}}^{\mathrm{orb}}(0)$ (T) & 8.18 & 9.1 & 8.25 \\
$B_{\mathrm{c2}}(0)$ (T) [$\alpha \sim 0$; WHH fit] & 8.12 & 8.9 & --- \\
$B_{\mathrm{c2}}(0)$ (T) [GL fit] & 10 & 10.5 & 13 \\
\end{tabular}
\end{ruledtabular}
\end{table}

The extracted parameters from the WHH and phenomenological Ginzburg--Landau (GL) analyses~\cite{tinkham2004introduction} are summarized in Table~\ref{tab:parameters}. For the as-grown films (A8 and A20), the experimental $B_{\mathrm{c2}}(T)$ data are well described within the conventional orbital-limited framework by setting $\alpha \sim 0$. As shown in Figs.~\ref{transitions}(d,e), the extrapolated zero-temperature upper critical fields $B_{\mathrm{c2}}(0)$ from the WHH model yield $8.12\text{ T}$ for A8 and $8.9\text{ T}$ for A20, which closely correspond to their respective orbital limits $B_{\mathrm{c2}}^{\mathrm{orb}}(0)$ and remain well below their Pauli limits $B_p(0)$. This conventional behavior indicates that orbital depairing dominates the upper critical field in the as-grown state, consistent with a predominantly spin-singlet superconducting state where paramagnetic effects play a minor role~\cite{werthamer1966temperature}. For completeness, phenomenological GL fits with $B_{\mathrm{c2}}(T) = B_{\mathrm{c2}}(0){(1-t^2) / (1+t^2)}$ yield $B_{\mathrm{c2}}(0) \approx 10\text{ T}$ for A8 and $10.5\text{ T}$ for A20, which remain below or close to the corresponding Pauli paramagnetic limits, confirming that the upper critical field of the as-grown films is largely governed by conventional pair-breaking mechanisms. 

By contrast, the annealed sample N20-30min exhibits a pronounced deviation from this conventional picture. As seen in Fig.~\ref{transitions}(f), the experimental $B_{\mathrm{c2}}(T)$ data systematically lie above the purely orbital-limited $\alpha = 0$ WHH prediction. Furthermore, a phenomenological GL analysis yields an extrapolated zero-temperature field $B_{\mathrm{c2}}(0) \approx 13\text{ T}$, which simultaneously exceeds both the orbital-limiting field ($B_{\mathrm{c2}}^{\mathrm{orb}}(0) = 8.25\text{ T}$) and the Pauli paramagnetic limit ($B_\mathrm{p}(0) = 11.6\text{ T}$). This enhancement of $B_{\mathrm{c2}}$ beyond the conventional theoretical boundaries cannot be captured by standard pair-breaking contributions; instead, it points to a substantial mitigation of paramagnetic pair breaking. 

In non-centrosymmetric superconductors, such a violation of the Pauli limit is a hallmark of strong antisymmetric spin--orbit coupling (ASOC), which locks the electron spins and protects the Cooper pairs from Zeeman splitting, potentially enabling a mixing of spin-singlet and spin-triplet components in the SC order parameter. It is worth noting that while generic spin--orbit scattering ($\lambda_{\mathrm{so}}$) in conventional, centrosymmetric WHH theory can only partially mitigate paramagnetic limiting---with the purely orbital-limited curve ($\alpha = 0$) representing an upper boundary that cannot be surpassed---the situation is fundamentally different in NCS systems. Here, the ASOC introduces a strong spin-momentum locking that acts as a robust internal field, directly shifting or completely suppressing the paramagnetic pair-breaking channel. Combined with the potential mixing of spin-singlet and spin-triplet components, this enables the experimental $B_{\mathrm{c2}}(T)$ to break past the conventional $\alpha = 0$ orbital ceiling, as observed in our annealed sample.

Finally, the $B_{\mathrm{c2}}(T)$ curves were used to estimate the normal-state electron diffusion coefficient $D = -1.097 \left(dB_{\mathrm{c2}}/dT\right)^{-1}|_{T=T_\mathrm{c}}$~\cite{werthamer1966temperature} and the zero-temperature coherence length $\xi(0) = \sqrt{\hbar D / 1.76 k_\mathrm{B} T_\mathrm{c}}$~\cite{tinkham2004introduction}.  For the as-grown films, $D$ decreases from $0.55\text{ cm}^2/\text{s}$ (A20) to $0.49\text{ cm}^2/\text{s}$ (A8), reflecting enhanced disorder and scattering in the thinner film. Interestingly, $D$ remains largely unchanged after annealing ($0.57\text{ cm}^2/\text{s}$ for N20-30min), suggesting that the overall electronic diffusivity is preserved despite the significant microstructural evolution. This indicates that while structural annealing reduces grain-boundary density, it may introduce compensating scattering centers, such as localized oxidation at the remaining boundaries. Notably, all investigated samples exhibit a nearly identical superconducting coherence length of $\xi(0) \approx 6\text{ nm}$.

\section{Normal-state magnetotransport}

\subsection{Phenomenological modeling}

To further elucidate the influence of thermal annealing on the electronic transport properties of Nb$_{0.18}$Re$_{0.82}$ thin films, we investigated their magnetotransport behavior in the normal state. Magnetoresistance (MR) measurements provide a sensitive probe of charge-carrier scattering, disorder, quantum interference, and spin--orbit coupling, thereby complementing the structural and superconducting analyses discussed in the previous sections.

Figure~\ref{magneto3}(a) presents the normalized magnetoresistance data $\Delta\rho(B)/\rho_{0}$, where $\Delta\rho(B)=\rho(B)-\rho_{0}$, $\rho(B)$ is the field-dependent resistivity, $\rho_{0}=\rho(0)$ is the zero-field resistivity, and $B$ is the applied magnetic field for the as-grown (A150) and annealed (N150) films measured at $T=8$\,K. Magnetotransport measurements were performed in a standard four-probe van der Pauw configuration by sweeping the magnetic field from $0$ to $7$\,T perpendicular to the film plane. The voltage was measured across the transverse voltage contacts, from which the longitudinal magnetoresistance was determined. 
To initially characterize the overall magnetic-field dependence and curvature of the WAL response without introducing \emph{a priori} microscopic parameters, we first employ the phenomenological description commonly used for nonlinear magnetotransport. This approach has previously been applied to investigate the MR of bulk and polycrystalline NbRe~\cite{sundar2019study}. The experimental data were fitted using the phenomenological relation~\cite{ziman1972principles}:
\begin{equation}
    \label{eq:MR}
    \frac{\Delta\rho(B)}{\rho_{0}}=\frac{B^{2}}{\alpha+\beta B^{2}}
\end{equation}

The fitting parameters $\alpha$ and $\beta$ describe the low-field curvature and high-field saturation behavior of the MR, respectively, providing a smooth interpolation between the quadratic low-field regime ($\Delta\rho(B)/\rho_{0}\approx B^{2}/\alpha$) and the saturation regime at high magnetic fields ($\Delta\rho(B)/\rho_{0}\rightarrow1/\beta$).

The results of the fitting procedure are shown as blu lines in Fig. ~\ref{magneto3}(a). The best-fit parameters obtained from Eq.~\eqref{eq:MR} are $\alpha=531\,\mathrm{T}^{2}$ and $\beta=908$ for the as-grown film A150, and $\alpha=77\,\mathrm{T}^{2}$ and $\beta=118$ for the annealed film N150. Here, $\alpha$ defines the characteristic magnetic-field scale of the MR response, while $\beta$ determines the saturation amplitude. The substantially smaller values of both $\alpha$ and $\beta$ obtained for N150 quantitatively reflect the significantly stronger MR response after thermal annealing. In particular, the reduced $\alpha$ corresponds to the enhanced low-field curvature, whereas the lower $\beta$ reflects the larger saturation amplitude. Such changes in the field-dependent transport behavior are consistent with enhanced quantum-interference effects, qualitatively suggesting an evolution of the transport properties and spin--orbit-related effects upon annealing.

However, although this empirical description successfully reproduces the overall field dependence of the MR, it remains purely phenomenological. It cannot distinguish the individual contributions of phase coherence and spin--orbit scattering, nor does it provide quantitative access to the intrinsic microscopic length scales governing electron transport. Furthermore, because thermal annealing may introduce additional scattering channels associated with oxidized grain boundaries and microstructural evolution, a more rigorous microscopic framework is required to elucidate the underlying transport mechanisms.

\subsection{Microscopic Kawabata 3D modeling}

\begin{figure}[t]
  \centering
  \includegraphics[width=0.8\columnwidth]{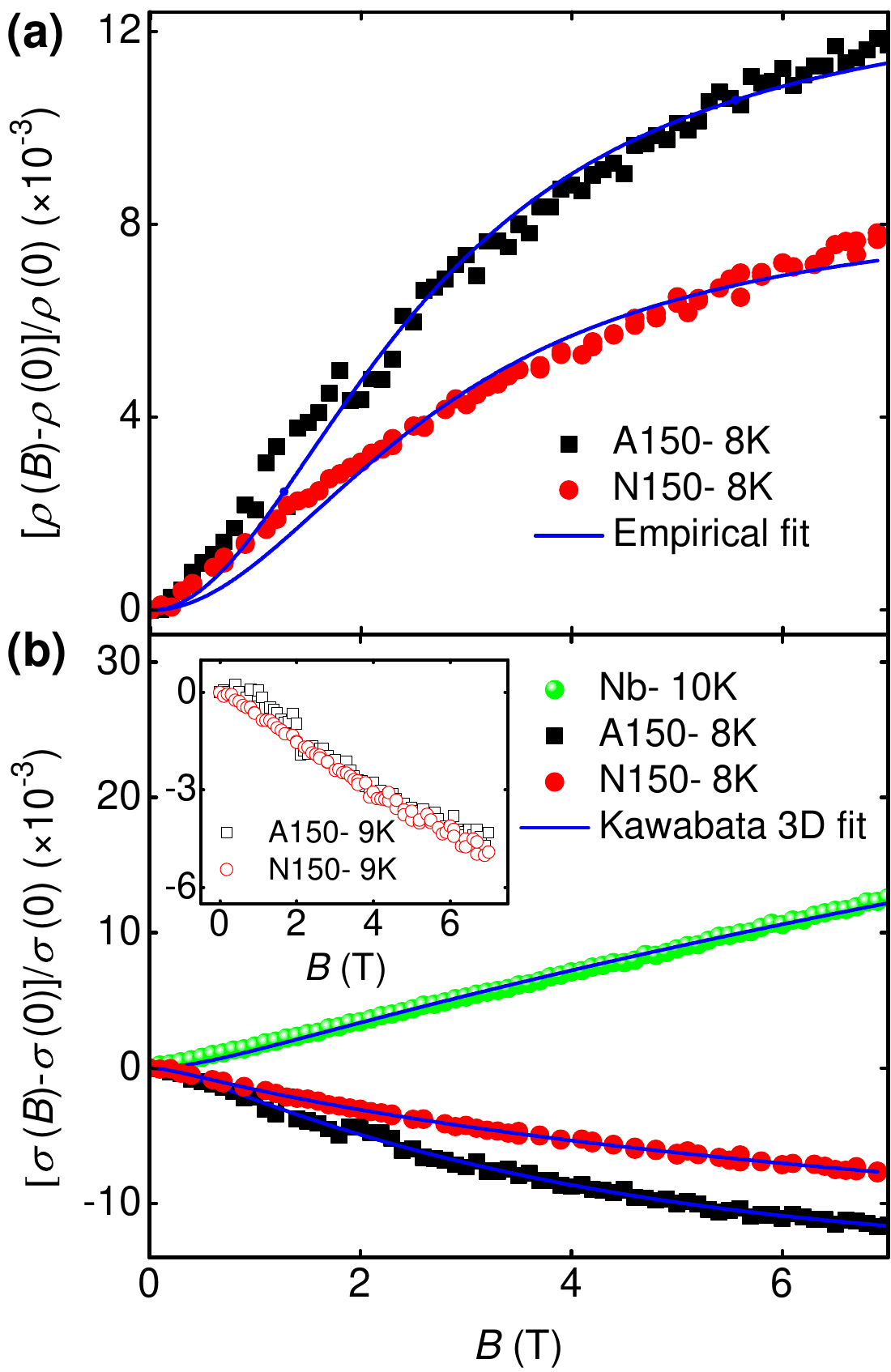}
  \caption{\label{magneto3}
    (a) Normalized magnetoresistance, $[\rho(B)-\rho(0)]/\rho(0)$, of the as-grown (A150) and annealed (N150) NbRe thin films measured in the normal state at $T = 8\text{ K}$. Solid lines represent fits to the phenomenological relation $\Delta\rho(B)/\rho_0 = B^2 / (\alpha + \beta B^2)$. 
    (b) Normalized magnetoconductivity, $[\sigma(B)-\sigma(0)]/\sigma(0)$, for the pure Nb reference film measured at $T = 10\text{ K}$, and for the A150 and N150 NbRe films measured at $T = 8\text{ K}$. Solid lines indicate the microscopic fits obtained using the 3D Kawabata weak-localization/weak-anti-localization framework. Inset: magnetoconductivity of the A150 and N150 films measured at $T=9$\,K, illustrating the reduced difference between the two samples at a temperature far above $T_\mathrm{c}$.}
\end{figure}

Figure~\ref{magneto3}(b) presents the MC of the A150 and N150 films measured in the normal state at $T=8$\,K over the field range $0$--$7$\,T. For comparison, the MC of a $150$\,nm-thick Nb reference film was measured at $T=10$\,K, above its $T_\mathrm{c}$. The data were analyzed within the 3D Kawabata WL/WAL framework~\cite{kawabata1980theory}, with the resulting fits shown in Fig.~\ref{magneto3}(b). The normalized MC correction,
$g(B)=\Delta\sigma(B)/\sigma_0=[\sigma(B)-\sigma(0)]/\sigma(0)$,
is described by the three-dimensional Kawabata WL/WAL model~\cite{kawabata1980theory},

\begin{equation}
\label{eq:app_kawabata_wlwal}
\begin{aligned}
g_{\mathrm{K}}(B)=
A_{\mathrm{K}}\sqrt{|B|}
\Bigg[
&\frac{3}{2}
f_3\!\left(
\frac{B_\phi+\frac{4}{3}B_{\mathrm{so}}}{|B|}
\right)
-
\frac{1}{2}
f_3\!\left(
\frac{B_\phi}{|B|}
\right)
\Bigg],
\end{aligned}
\end{equation}
where $A_{\mathrm{K}}$ is a fitting prefactor describing the amplitude of the quantum-interference correction, $f_3(x)$ is the Kawabata function 
\begin{equation}
\label{eq:Kawabata_func}
f_3(x)=
\sum_{n=0}^{\infty}
\left[
\frac{2}{\sqrt{n+1+x}+\sqrt{n+x}}
-
\frac{1}{\sqrt{n+\frac12+x}}
\right].
\end{equation}
and $B_\phi$ and $B_{\mathrm{so}}$ are the dephasing and spin--orbit scattering fields, respectively. These characteristic fields are related to the phase-coherence length $L_\phi$ and spin--orbit scattering length $L_{\mathrm{so}}$ through

\begin{equation}
\label{eq:Bfields}
B_\phi=\frac{\hbar}{4eL_\phi^2},
\qquad
B_{\mathrm{so}}=\frac{\hbar}{4eL_{\mathrm{so}}^2}.
\end{equation}

Physically, $B_\phi$ defines the magnetic-field scale over which quantum interference is suppressed, whereas $B_{\mathrm{so}}$ quantifies the strength of spin--orbit scattering. The corresponding length scales are given by

\begin{equation}
\label{eq:lengthscales}
L_\phi[\mathrm{nm}]
=
\frac{12.83}{\sqrt{B_\phi[\mathrm{T}]}},
\qquad
L_{\mathrm{so}}[\mathrm{nm}]
=
\frac{12.83}{\sqrt{B_{\mathrm{so}}[\mathrm{T}]}}.
\end{equation}

In the weak-SOC limit, $B_{\mathrm{so}}\ll B_\phi$, Eq.~\eqref{eq:app_kawabata_wlwal} reduces to
\begin{equation}
    \label{eq:app_fit_nb}
    g_{\mathrm{fit}}^{\mathrm{Nb}}(B)
    =
    A_{\mathrm{K}}\sqrt{|B|}
    f_3\!\left(\frac{B_\phi}{|B|}\right),
\end{equation}
which describes the positive MC characteristic of WL. The Nb reference film is well described by Eq.~\eqref{eq:app_fit_nb}, and the positive conductivity correction observed over the entire field range is consistent with weak spin--orbit scattering and WL-dominated transport, as shown in Fig.~\ref{magneto3}(b).

By contrast, the MC of both NbRe films is negative, consistent with the strong-SOC regime ($B_{\mathrm{so}}\gg B_\phi$), where the WAL contribution dominates the quantum-interference correction. In this regime, spin--orbit scattering changes the sign of the interference correction, suppressing backscattering and increasing the zero-field conductivity. Application of a magnetic field progressively suppresses the WAL contribution, leading to a decrease in conductivity with increasing field.

The fitting parameters extracted from the Kawabata analysis are summarized in Table~\ref{tab:kawabata_fit}. For all samples, the phase-coherence length remains substantially smaller than the film thickness ($L_\phi\sim20$--$40$\,nm $\ll d=150$\,nm), confirming that the quantum-interference correction is governed by 3D diffusive transport and supporting the applicability of the 3D Kawabata formalism rather than the 2D Hikami--Larkin--Nagaoka (HLN) WL model~\cite{hikami1980spin,bergmann1984weak}. Furthermore, both NbRe films satisfy the condition $B_{\mathrm{so}}\gg B_\phi$, or equivalently $L_{\mathrm{so}}\ll L_\phi$, placing them in the WAL regime.

\begin{table}[t!]
\caption{Extracted fitting parameters from the 3D Kawabata WL/WAL analysis.}
\label{tab:kawabata_fit}
\centering
\begin{tabular}{lcccccc}
\hline\hline
Sample & $T$(K) & $A_{\mathrm K}$ & $B_\phi$(T) & $L_\phi$(nm) & $B_{\mathrm{so}}$(T) & $L_{\mathrm{so}}$(nm) \\
\hline
Nb150 & 10 & $1.56\times10^{-2}$ & 0.281 & 24.2 & \text{--} & \text{--} \\
N150  & 8  & $1.57\times10^{-2}$ & 0.090 & 42.7 & 7.36 & 4.73 \\
A150  & 8  & $3.43\times10^{-2}$ & 0.155 & 32.6 & 3.25 & 7.12 \\
\hline\hline
\end{tabular}
\end{table}

A more quantitative measure of the relative spin--orbit scattering strength is provided by the ratio $B_{\mathrm{so}}/B_\phi$. At $T=8$\,K, this ratio increases from approximately 20 in the as-grown film to nearly 80 in the annealed film, indicating a substantial enhancement of spin--orbit scattering relative to dephasing after annealing. The stronger spin--orbit scattering in N150 is consistent with the enhanced unconventional features inferred from the $B_\mathrm{c2}(T)$ analysis and the deviation from conventional Pauli-limited behavior discussed above. It is also compatible with previous structural studies linking annealing to improved crystallinity and the emergence of the non-centrosymmetric $\chi$ phase in NbRe~\cite{makhdoumi2024effect}.

As shown in the inset of Fig.~\ref{magneto3}(b), the difference between the MC of A150 and N150 is pronounced at $8$\,K but becomes much weaker at $9$\,K. This behavior is consistent with the reduced magnitude of quantum-interference corrections at elevated temperatures, together with the diminishing influence of SC fluctuation effects as the temperature moves further above $T_\mathrm{c}$. Consequently, the MC becomes increasingly dominated by the normal-state background, reducing the distinction between the two films.

Finally, we emphasize that the present MC analysis does not provide direct evidence for spin-triplet SC. Rather, it demonstrates enhanced spin--orbit scattering, which can enable singlet--triplet mixing in NCSs.

\section{Conclusions}

In conclusion, we have systematically investigated the impact of thermal annealing on the superconducting and normal-state transport properties of non-centrosymmetric $\text{Nb}_{0.18}\text{Re}_{0.82}$ thin films. Although the thermal treatment induces a minor suppression of the zero-field transition temperature, it profoundly reorganizes the microstructural properties, altering the electronic response to high magnetic fields. 

In the superconducting state, the upper critical field of as-grown films remains well within the conventional orbital-limiting framework. Conversely, the annealed counterpart exhibits a distinct violation of the Pauli paramagnetic limit, indicating a substantial mitigation of paramagnetic pair-breaking. In non-centrosymmetric systems, this behavior serves as a clear hallmark of strong antisymmetric spin--orbit coupling, which stabilizes Cooper pairs against Zeeman splitting and favors the potential admixture of singlet and triplet components in the order parameter.

This mechanism has been independently validated through normal-state magnetotransport measurements. The experimental magnetoresistance demonstrates a sharp weak anti-localization  profile driven by the heavy Re atoms. Microscopic modeling within the three-dimensional Kawabata framework reveals a significant shortening of the spin--orbit scattering length $L_{\mathrm{so}}$ upon annealing. Crucially, since the electron diffusion coefficient $D$ remains nearly invariant, this variation cannot be ascribed to a simple reduction of the global mean free path; instead, it provides direct proof of a local enhancement of the spin--orbit interaction strength.

The convergence of the superconducting boundary analysis and the normal-state weak anti-localization modeling confirms a consistent physical picture where the stabilization of the non-centrosymmetric state under high fields is explicitly driven by the amplification of spin--orbit-mediated quantum transport. Overall, our findings establish thermal annealing as a powerful tuning parameter to engineer spin--orbit effects in transition-metal-based alloy films, offering promising pathways for designing quantum materials tailored for superconducting spintronics.

\section*{Data availability} 
The data that support the findings of this study are available from the corresponding author upon request.

\section*{Acknowledgements}
The work of Z.M.K. was funded by the Deutsche Forschungsgemeinschaft (DFG, German Research Foundation) under Grant No. 583926508 (SyMPaThY). Z.M.K. gratefully acknowledges support from the Braunschweig International Graduate School of Metrology (B-IGSM). Y.Y. acknowledges CryoQuant/TU Braunschweig for hospitality and financial support from the Deutsche Forschungsgemeinschaft (DFG, German Research Foundation) under Germany's Excellence Strategy – EXC-2123 QuantumFrontiers – 390837967. O.D. acknowledges support from the DFG through Grant No. 577881064 (Super3DMag). This research is based on work from COST Actions CA21144 (SuperQuMap) and CA23134 (Polytopo), supported by the European Cooperation in Science and Technology. This research was partially supported by the project
IR0000003–IRIS, funded by the NextGeneration EU-funded
Italian National Recovery and Resilience Plan with the Decree
of the Ministry of University and Research number 124 (21/06/
2022) for Mission 4 – Component 2 – Investment 3.1. This
research was also partially supported by the project “High-performance
Josephson junctions for ferrotransmons –
CONJUNCTIONS” in the frame of Partenariato Esteso “NQSTI”,
Spoke 5, funded by the Italian National Recovery and
Resilience Plan Mission 4, Component 2, Investment 3.1.

\bibliography{references}

@article{Smidman2017,
  author  = {M. Smidman and M. B. Salamon and H. Q. Yuan and D. F. Agterberg},
  title   = {Superconductivity and spin--orbit coupling in non‑centrosymmetric materials},
  journal = {Rep. Prog. Phys.},
  volume  = {80},
  number  = {3},
  pages   = {036501},
  year    = {2017},
  doi     = {10.1088/1361-6633/80/3/036501}
}

@article{Cirillo2016,
  author  = {C. Cirillo and G. Carapella and M. Salvato and R. Arpaia and M. Caputo and C. Attanasio},
  title   = {Superconducting properties of noncentrosymmetric {Nb}$_{0.18}${Re}$_{0.82}$ thin films probed by transport and tunneling experiments},
  journal = {Phys. Rev. B},
  volume  = {94},
  pages   = {104512},
  year    = {2016},
  doi     = {10.1103/PhysRevB.94.104512}
}

@article{Cirillo2015,
  author  = {C. Cirillo and R. Fittipaldi and M. Smidman and G. Carapella and C. Attanasio and A. Vecchione and R. P. Singh and M. R. Lees and G. Balakrishnan and M. Cuoco},
  title   = {Evidence of double‑gap superconductivity in noncentrosymmetric {Nb}$_{0.18}${Re}$_{0.82}$ single crystals},
  journal = {Phys. Rev. B},
  volume  = {91},
  pages   = {134508},
  year    = {2015},
  doi     = {10.1103/PhysRevB.91.134508}
}

@article{Karki2011,
  author  = {A. B. Karki and Y. M. Xiong and N. Haldolaarachchige and S. Stadler and I. Vekhter and P. W. Adams and D. P. Young and W. A. Phelan and J. Y. Chan},
  title   = {Physical properties of the noncentrosymmetric superconductor {Nb}$_{0.18}${Re}$_{0.82}$},
  journal = {Phys. Rev. B},
  volume  = {83},
  pages   = {144525},
  year    = {2011},
  doi     = {10.1103/PhysRevB.83.144525}
}

@article{Chen2013,
  title={{BCS}-like superconductivity in the noncentrosymmetric compounds {Nb}$_x${Re}$_{1-x}$ (0.13 $\leq x \leq$ 0.38)},
  author={J. Chen and L. Jiao and J. L. Zhang and Y. Chen and L. Yang and M. Nicklas and F. Steglich and H. Q. Yuan},
  journal={Phys. Rev. B},
  volume={88},
  pages={144510},
  year={2013},
  doi={10.1103/PhysRevB.88.144510}
}

@article{cirillo2022polycrystalline,
  title={Polycrystalline {NbRe} superconducting films deposited by direct current magnetron sputtering},
  author={C. Cirillo and M. Caputo and G. Divitini and J. W. A. Robinson and C. Attanasio},
  journal={Thin Solid Films},
  volume={758},
  pages={139450},
  year={2022},
  doi={10.1016/j.tsf.2022.139450}
}

@article{makhdoumi2024effect,
  title={Effect of thermal annealing on the average and local structure of superconducting polycrystalline {NbRe} films},
  author={{Z. Makhdoumi Kakhaki} and A. Martinelli and F. Avitabile and A. Di Bernardo and C. Attanasio and C. Cirillo},
  journal={Supercond. Sci. Technol.},
  volume={37},
  number={12},
  pages={125002},
  year={2024},
  doi={10.1088/1361-6668/ad8122}
}

@article{Linder2015,
  author  = {J. Linder and J. W. A. Robinson},
  title   = {Superconducting spintronics},
  journal = {Nat. Phys.},
  volume  = {11},
  pages   = {307},
  year    = {2015},
  doi     = {10.1038/nphys3242}
}

@article{kakhaki2023upper,
  author  = {{Z. Makhdoumi Kakhaki} and A. Leo and F. Chianese and L. Parlato and G. P. Pepe and A. Nigro and C. Cirillo and C. Attanasio},
  title   = {Upper critical magnetic field in {NbRe} and {NbReN} micrometric strips},
  journal = {Beilstein J. Nanotechnol.},
  volume  = {14},
  pages   = {45--51},
  year    = {2023},
  doi     = {10.3762/bjnano.14.5}
}

@article{caputo2017nbre,
  author  = {M. Caputo and C. Cirillo and C. Attanasio},
  title   = {{NbRe} as candidate material for fast single photon detection},
  journal = {Appl. Phys. Lett.},
  volume  = {111},
  pages   = {192601},
  year    = {2017},
  doi     = {10.1063/1.4997675}
}

@article{Maki1964,
  author  = {K. Maki},
  title   = {The magnetic properties of superconducting alloys. {II}},
  journal = {Phys. Phys. Fiz.},
  volume  = {1},
  pages   = {127--136},
  year    = {1964}
}

@book{ziman1972principles,
  author    = {J. M. Ziman},
  title     = {{Principles of the Theory of Solids}},
  publisher = {Cambridge Univ. Press},
  year      = {1972}
}

@article{cirillo2020superconducting,
  author  = {C. Cirillo and J. Chang and M. Caputo and J. W. N. Los and S. Dorenbos and I. Esmaeil Zadeh and C. Attanasio},
  title   = {Superconducting nanowire single photon detectors based on disordered {NbRe} films},
  journal = {Appl. Phys. Lett.},
  volume  = {117},
  pages   = {172602},
  year    = {2020},
  doi     = {10.1063/5.0021487}
}

@article{ejrnaes2022single,
  author  = {M. Ejrnaes and C. Cirillo and D. Salvoni and F. Chianese and C. Bruscino and P. Ercolano and A. Cassinese and C. Attanasio and G. P. Pepe and L. Parlato},
  title   = {Single photon detection in {NbRe} superconducting microstrips},
  journal = {Appl. Phys. Lett.},
  volume  = {121},
  pages   = {265303},
  year    = {2022},
  doi     = {10.1063/5.0131336}
}

@article{cirillo2024single,
  author  = {C. Cirillo and M. Ejrnaes and P. Ercolano and C. Bruscino and A. Cassinese and D. Salvoni and C. Attanasio and G. P. Pepe and L. Parlato},
  title   = {Single photon detection up to 2 \(\mu\)m in pair of parallel microstrips based on {NbRe} ultrathin films},
  journal = {Sci. Rep.},
  volume  = {14},
  pages   = {20345},
  year    = {2024},
  doi     = {10.1038/s41598-024-66991-1}
}

@article{koch2024gate,
  author  = {J. Koch and C. Cirillo and S. Battisti and L. Ruf and {Z. Makhdoumi Kakhaki} and A. Paghi and A. Gulian and S. Teknowijoyo and G. De Simoni and F. Giazotto and C. Attanasio and E. Scheer and A. {Di Bernardo}},
  title   = {Gate‑controlled supercurrent effect in dry‑etched {Dayem} bridges of non‑centrosymmetric niobium rhenium},
  journal = {Nano Res.},
  volume  = {17},
  pages   = {6575},
  year    = {2024},
  doi     = {10.1007/s12274-024-6576-7}
}

@article{colangelo2025unveiling,
  author  = {F. Colangelo and M. Modestino and F. Avitabile and A. Galluzzi and Z. Makhdoumi Kakhaki and A. Kumar and J. Linder and M. Polichetti and C. Attanasio and C. Cirillo},
  title   = {Unveiling Intrinsic Triplet Superconductivity in Noncentrosymmetric {NbRe} through Inverse Spin-Valve Effects},
  journal = {Phys. Rev. Lett.},
  volume  = {135},
  number  = {22},
  pages   = {226002},
  year    = {2025},
  doi     = {10.1103/q1nb-cvh6}
}

@article{sundar2019study,
  author  = {S. Sundar and S. Salem‑Sugui and M.~K. Chattopadhyay and S.~B. Roy and L.~S. Sharath Chandra and L.~F. Cohen and L. Ghivelder},
  title   = {Study of {Nb}$_{0.18}${Re}$_{0.82}$ non‑centrosymmetric superconductor in the normal and superconducting states},
  journal = {Supercond. Sci. Technol.},
  volume  = {32},
  number  = {5},
  pages   = {055003},
  year    = {2019},
  doi     = {10.1088/1361-6668/ab06a5}
}

@article{kakhaki2024characterization,
  author  = {{Z. Makhdoumi Kakhaki} and A. Leo and A. Spuri and M. Ejrnaes and L. Parlato and G. P. Pepe and F. Avitabile and A. {Di Bernardo} and A. Nigro and C. Attanasio and C. Cirillo},
  title   = {Characterization of quasiparticle relaxation times in microstrips of {NbReN} for prospective applications in superconducting single‑photon detectors},
  journal = {Mater. Sci. Eng. B},
  volume  = {304},
  pages   = {117376},
  year    = {2024},
  doi     = {10.1016/j.mseb.2024.117376}
}

@book{bauer2012non,
  author    = {E. Bauer and M. Sigrist},
  title     = {{Non-Centrosymmetric Superconductors}: Introduction and Overview},
  series    = {Lecture Notes in Physics},
  volume    = {847},
  publisher = {Springer},
  year      = {2012},
  doi       = {10.1007/978-3-642-24624-1}
}

@article{de2026fast,
  title={Fast vortex dynamics and relaxation times in {NbRe}-based heterostructures},
  author={F. {De Chiara} and Z. {Makhdoumi Kakhaki} and F. Avitabile and F. Colangelo and A. Kumar and C. Attanasio and C. Cirillo},
  journal={Beilstein J. Nanotechnol.},
  volume={17},
  number={1},
  pages={292},
  year={2026},
  doi={10.3762/bjnano.17.20}
}

@article{battisti2024demonstration,
  title={Demonstration of high-impedance superconducting {NbRe Dayem} bridges},
  author={S. Battisti and J. Koch and A. Paghi and L. Ruf and A. Gulian and S. Teknowijoyo and C. Cirillo and {Z. Makhdoumi Kakhaki} and C. Attanasio and E. Scheer and A. {Di Bernardo} and G. De Simoni and F. Giazotto},
  journal={Appl. Phys. Lett.},
  volume={124},
  number={17},
  pages={172601},
  year={2024},
  doi={10.1063/5.0200257}
}

@book{tinkham2004introduction,
  title={Introduction to Superconductivity},
  author={Michael Tinkham},
  publisher={Dover Publications},
  address={Mineola, New York},
  year={2004},
  edition={2},
  doi={10.1063/1.2807811}
}

@article{eschrig2011spin,
  author  = {Eschrig, Matthias},
  title   = {Spin-polarized supercurrents for spintronics},
  journal = {Physics Today},
  volume  = {64},
  number  = {1},
  pages   = {43},
  year    = {2011},
  doi     = {10.1063/1.3541944}
}

@article{khaire2010observation,
  author  = {Khaire, Trupti S. and Khasawneh, Mazin A. and Pratt Jr., W. P. and Birge, Norman O.},
  title   = {Observation of spin-triplet superconductivity in {Co}-based {J}osephson Junctions},
  journal = {Phys. Rev. Lett.},
  volume  = {104},
  pages   = {137002},
  year    = {2010},
  doi     = {10.1103/PhysRevLett.104.137002}
}

@article{robinson2010controlled,
  author  = {Robinson, J. W. A. and Witt, J. D. S. and Blamire, M. G.},
  title   = {Controlled injection of spin-triplet supercurrents into a strong ferromagnet},
  journal = {Science},
  volume  = {329},
  pages   = {59--61},
  year    = {2010},
  doi     = {10.1126/science.1189246}
}

@article{ShangPRL2018,
  author  = {T. Shang and M. Smidman and S. K. Ghosh and C. Baines and L. J. Chang and D. J. Gawryluk and J. A. T. Barker and R. P. Singh and D. McK. Paul and G. Balakrishnan and E. Pomjakushina and M. Shi and M. Medarde and A. D. Hillier and H. Q. Yuan and J. Quintanilla and J. Mesot and T. Shiroka},
  title   = {Time-Reversal Symmetry Breaking in {Re}-Based Superconductors},
  journal = {Phys. Rev. Lett.},
  volume  = {121},
  pages   = {257002},
  year    = {2018},
  doi     = {10.1103/PhysRevLett.121.257002}
}

@article{Dob25nan,
  author = {Dobrovolskiy, O. V. and Wang, Q. and Vodolazov, D. Yu. and Sachser, R. and Huth, M. and Knauer, S. and Buzdin, A. I.},
  title = {Moving {Abrikosov} vortex lattices generate sub-40-nm magnons},
  journal = {Nat. Nanotechn.},
  year = {2025},
  volume = {20},
  number = {12},
  pages = {1764--1770},
  doi = {10.1038/s41565-025-02024-w}
}

@article{cirillo2023spin,
  title   = {Spin pumping in {NbRe/Co} superconductor-ferromagnet heterostructures},
  author  = {Cirillo, Carla and Rovirola, Marc and Gonz{\'a}lez, Carla and Casals, Blai and Hern{\`a}ndez, Joan Manel and Maci{\`a}, Ferran and Garc{\'\i}a-Santiago, Antoni and Attanasio, Carmine},
  journal = {Supercond. Sci. and Technol.},
  volume  = {36},
  number  = {7},
  pages   = {074001},
  year    = {2023},
  publisher = {IOP Publishing},
  doi     = {10.1088/1361-6668/acd1c0}
}

@article{werthamer1966temperature,
  title   = {Temperature and Purity Dependence of the Superconducting Critical Field, {$H_{c2}$. III.} Electron Spin and Spin-Orbit Effects},
  author  = {Werthamer, N. R. and Helfand, Eugene and Hohenberg, Pierre C.},
  journal = {Phys. Rev.},
  volume  = {147},
  number  = {1},
  pages   = {295--302},
  year    = {1966},
  publisher = {American Physical Society},
  doi     = {10.1103/PhysRev.147.295}
}

@article{bergmann1984weak,
  author  = {Bergmann, Gerd},
  title   = {Weak localization in thin films: a time-of-flight experiment with conduction electrons},
  journal = {Phys. Rep.},
  volume  = {107},
  number  = {1},
  pages   = {1--58},
  year    = {1984},
  doi     = {10.1016/0370-1573(84)90103-0}
}

@article{hikami1980spin,
  author  = {S. Hikami and A. I. Larkin and Y. Nagaoka},
  title   = {Spin-Orbit Interaction and Magnetoresistance in the Two-Dimensional Random System},
  journal = {Prog. Theor. Phys.},
  volume  = {63},
  number  = {2},
  pages   = {707--710},
  year    = {1980},
  doi     = {10.1143/PTP.63.707}
}

@article{kawabata1980theory,
  title={Theory of negative magnetoresistance. {I. Application to heavily doped semiconductors}},
  author={A. Kawabata},
  journal={J. Phys. Soc. Jpn.},
  volume={49},
  number={2},
  pages={628--637},
  year={1980},
  doi={10.1143/JPSJ.49.628}
}

@ARTICLE{Dob26sst,
  author = {Dobrovolskiy, O. and Suderow, H. and Tafuri, F. and Black-Schaffer,
	A. M. and Lado, J. L. and Sudb\o{}, A. and Stornaioulo, D. and Li,
	C. and B\"ohmer, A. E. and Tran, L. M. and Zaleski, A. J. and Crisan,
	A. and Polichetti, M. and Galluzzi, A. and Gencer, A. and Aichner,
	B. and Bari\'si\'c, N. and Lang, W. and Samuely, T. and Gmitra, M.
	and Cren, T. and Calandra, M. and Samuely, P. and Custers, J. and
	C\'ordoba, R. and Fomin, V. M. and Poccia, N. and Szab\'o, P. and
	Porrati, F. and Kakazei, G. and Aarts, J. and Robinson, J. and Villegas,
	J. E. and Althammer, M. and Huebl, H. and Kamra, A. and Weiler, M.
	and Dil, J. H. and Evtushinsky, D. and Kalisky, B. and Anahory, Y.
	and Bending, S. and Liljeroth, P. and Hassanien, A. and Guillam\'on,
	I. and Herrera, E. and Silhanek, A. V. and Van de Vondel, J. and
	Palau, A. and Charaev, I. and Sidorova, M. and Lombardi, F. and Bauch,
	T. and Feuillet-Palma, C. and Stolyarov, V. and Roditchev, D. and
	Krasnov, V. M. and Hampel, B. and Mart\'{\i}nez-P\'erez, M. J. and
	Ses\'e, J. and Koelle, D. and Poletto, S. and Bruno, A. and Massarotti,
	D.
	
	},
  title = {Roadmap on nanoscale superconductivity for quantum technologies},
  journal = {Supercond. Sci. Technol.},
  year = {2026},
  volume = {39},
  pages = {023502},
  number = {2},
  doi = {10.1088/1361-6668/ae3030},
  publisher = {IOP Publishing},
  url = {https://doi.org/10.1088/1361-6668/ae3030}
}

@article{Yip_review,
   author = "Yip, Sungkit",
   title = "Noncentrosymmetric Superconductors", 
   journal= "Annu. Rev. Condens. Matter Phys.",
   year = "2014",
   volume = "5",
   number = "Volume 5, 2014",
   pages = "15-33",
   doi = "https://doi.org/10.1146/annurev-conmatphys-031113-133912",
   url = "https://www.annualreviews.org/content/journals/10.1146/annurev-conmatphys-031113-133912",
   publisher = "Annual Reviews",
   issn = "1947-5462",
   type = "Journal Article",
  }

@article{grigorov1993interdiffusion,
  author  = {K. G. Grigorov and G. I. Grigorov and M. V. Stoyanova and R. A. Chakalov and J. L. Vignes and J. P. Langeron and P. Denjean and J. Perriere},
  title   = {Interdiffusion of {Y--Ba--Cu} oxides and {SiO}$_2$ substrate: Efficiency of {TiN} barrier film},
  journal = {Vacuum},
  volume  = {44},
  number  = {11--12},
  pages   = {1119--1121},
  year    = {1993},
  doi     = {10.1016/0042-207X(93)90235-K}
}

@article{piallat2016investigation,
  author  = {F. Piallat and R. Gassilloud and P. Caubet and C. Vallee},
  title   = {Investigation of {TiN} Thin Film Oxidation Depending on the Substrate Temperature at Vacuum Break},
  journal = {J. Vac. Sci. Technol. A},
  volume  = {34},
  number  = {5},
  pages   = {051508},
  year    = {2016},
  doi     = {10.1116/1.4960648}
}

\end{document}